\input prepictex
\input pictex
\input postpictex
\newcommand{\ptcommand}[1]{\put(0,0){\beginpicture
\setcoordinatesystem units <\unitlength,\unitlength>
#1
\endpicture}}

\newcommand{\ds}{\displaystyle}
\newcommand{\eqr}[1]{(\ref{#1})}

\newcommand{\be}{\begin{equation}}
\newcommand{\ee}{\end{equation}}
\newcommand{\bp}{\begin{picture}}
\newcommand{\ep}{\end{picture}}

\newcommand{\ba}[1]{\begin{array}{#1}\ds }
\newcommand{\cra}{\\ \ds}
\newcommand{\ea}{\end{array}}

\newcommand{\etc}{{\it etc.\/}}

\def\cL{{\cal L}}

\def\bea{\begin{eqnarray}}
\def\eea{\end{eqnarray}}

\documentstyle[ichep,12pt]{article}
\title{\normalsize RENORMALIZATION GROUP COEFFICIENTS\\
       FOR THE UN-TRUNCATED DERIVATIVE EXPANSION\\
        IN EFFECTIVE FIELD THEORY
       \thanks{Research supported in part by NSF Grant No. PHY-8714654 and
TNRLC Grant No. RGFY9106. }}
\author{Vineer Bhansali \\
        Lyman Labs. of Physics, Harvard University \\
       Cambridge, MA 02138, USA}

\begin{document}
\finalcopy

\maketitle

\abstract{
We investigate the renormalization of ``nonlocal" interactions
which arise as an infinite sum of higher derivative interactions in an
effective field theory. Using dimensional regularization with minimal
subtraction in a general scalar field theory, we write an
integro-differential renormalization group equation for the most general
graph at one loop order.}

\vskip-1pc
\onehead{INTRODUCTION}

In its traditional form, an effective field theory calculation goes like this:
Start at a very large scale, that is with the renormalization scale, $\mu$,
very large. In a strongly interacting theory or a theory with unknown physics
at high energy, this starting scale should be sufficiently large that
nonrenormalizable interactions produced at higher scales are too small to be
relevant. In a renormalizable, weakly interacting theory, one starts at a
scale above the masses of all the particles, where the effective theory is
given simply by the renormalizable theory, with no nonrenormalizable terms.
The theory is then evolved down to lower scales. As long as no particle masses
are encountered, this evolution is described by the renormalization group.
However, when $\mu$ goes below the mass, $\Lambda$, of one of the particles in
the theory, we must change the effective theory to a new theory without that
particle. In the process, the parameters of the theory
\break
\newpage\vglue1pt
\noindent
change, and new,
nonrenormalizable interactions may be introduced. Both the changes
in existing
parameters, and the coefficients of the new interactions are computed by
``matching'' the physics just below the boundary in the two theories. It is
this process that determines the relative sizes of the nonrenormalizable terms
associated with the heavy particles.
Because matching is done for $\mu\approx\Lambda$, the rule for the size of
the coefficients of the new operators is simple for $\mu\approx\Lambda$. At
this scale, all the new contributions scale with $\Lambda$ to the
appropriate power (set by dimensional analysis) up to factors of coupling
constants, group theory or counting factors and loop factors (of $16\pi^2$,
\etc).$^1$ Then when the new effective theory is evolved down to smaller
$\mu$, the renormalization group introduces additional factors into the
coefficients. Thus a heavy particle mass appears in the parameters of an
effective field theory in two ways. There is power dependence on the mass
that arises from matching conditions. There is also logarithmic dependence
that arises from the renormalization group.

The matching correction at tree level is simply a difference between a
calculation in the full theory and a calculation in the low energy effective
theory
\be\ba{c}
\int\delta\cL^0(\Phi)=S_{\cL_H+\cL}^0(\Phi)-S^0_{\cL}(\Phi)\cra
=\int{\textstyle\left\{
{\rm virtual\;heavy\atop particle\;trees}\right\}}(\Phi)\ea
\ee
where $S_{\cL_H+\cL}^0(\Phi)$ denotes the light particle effective action in
the full theory and $S^0_{\cL}(\Phi)$ denotes the same in the low energy
theory.$^2$
The matching correction so obtained is nonlocal because it depends on
$p/\Lambda$ through the virtual heavy particle propagators. It is also \bf
analytic \rm in $p/\Lambda$ in the region relevant to the low energy theory,
i.e. for characteristic momentum $<< \Lambda$. Thus it can be expanded in
powers of $p /\Lambda$ with the
higher order terms decreasing in importance: this corresponds to a local
operator
product expansion in the domain of analyticity, equivalent
to a local nonrenormalizable Lagrangian which can be treated as
an honest-to-goodness local field theory. However in general an infinite
series of terms of increasingly higher dimension are generated by matching at
tree level. These cause no problem when the scales are well separated, because
their effects quickly become negligible. But if there are two or more scales
close together, then we may not be justified
in ignoring terms at higher orders in the expansion.
(How) can we keep track of all the infinite number of higher derivative
operators efficiently?  In the
particular context of a scalar field theory, we
answer this question at the one loop level. The approach is direct. We show
that the resulting $\beta$ functions for the terms in the momentum
expansion can be collected into integro-differential renormalization group
equations for nonlocal couplings. The interested reader may refer to
ref. 3 for details of computations.

\break
\newpage
\onehead{GENERAL RESULTS}

Working within the formalism of a massive nonlocal effective
theory induced by some unknown full theory, we are forced to consider an
effective Lagrangian with operators with an arbitrary number of low energy
fields. In this section, we will compute the
one-loop running of a general non-renormalizable,
nonlocal scalar effective theory with $\phi\rightarrow-\phi$ symmetry.
Specifically, our purpose is to
isolate the dimensional regularization pole of a $2m$ point function of type
$n$ (i.e. with $n$ vertices or propagators) at one loop.
We give the complete
expression for the $1/ \epsilon$ pole of the $2m$ point function of type $n$
at one loop as a surprisingly compact
multi-Feynman-parameter, multi-dimensional angular integral.

We assume that the dimensionally continued $d=4-\epsilon$ dimensional
non-local Lagrangian has a $\phi\rightarrow-\phi$ symmetry and hence has an
interaction term proportional to
\be \cL_{\rm int.} = \sum_{r=1}^{\infty} {\mu}^{\epsilon(r-1)} G_{2r}
\phi^{2r}. \label{coupling} \ee
The mass scale $\mu$ is introduced to keep the dimensions of the nonlocal
couplings $G_{2r}$ fixed under dimensional continuation. The interaction is
not normal ordered, so fields at the same point can be contracted,
and tadpoles occur explicitly, and each $G_{2r}$ is some nonlocal
function which is analytic in the region under consideration, depends as a
consequence of momentum conservation on $2r-1$ linearly independent momenta,
and may have dimensions proportional to some power of an implicit scale of
nonlocality $\Lambda$.

\begin{figure}
$$\bp(200,160)(-100,-80)
\thicklines
\put(0,0){\ptcommand{\setplotsymbol ({\small .})
\circulararc 360 degrees from 60 0 center at 0 0}}
\put(0,-60){\circle*{8}}
\put(-60,0){\circle*{8}}
\put(60,0){\circle*{8}}
\put(42.4,42.4){\circle*{8}}
\put(-42.4,42.4){\circle*{8}}
\put(42.4,-42.4){\circle*{8}}
\put(-42.4,-42.4){\circle*{8}}
\put(-64.7,26.8){\makebox(0,0)[r]{$k$}}
\put(-64.7,-26.8){\makebox(0,0)[r]{$k+Q_1$}}
\put(-26.8,-64.7){\makebox(0,0)[rt]{$k+Q_2$}}
\put(26.8,-64.7){\makebox(0,0)[lt]{$k+Q_3$}}
\put(-26.8,64.7){\makebox(0,0)[rb]{$k+Q_{n-1}$}}
\put(64.7,-26.8){\makebox(0,0)[l]{$k+Q_4$}}
\put(64.7,26.8){\makebox(0,0)[lb]{$k+Q_5$}}
\put(26.8,64.7){\makebox(0,0)[lb]{$k+Q_{6}$}}
\put(-49.5,49.5){\makebox(0,0){$P_n$}}
\put(-49.5,-49.5){\makebox(0,0){$P_2$}}
\put(49.5,-49.5){\makebox(0,0){$P_4$}}
\put(49.5,49.5){\makebox(0,0){$P_6$}}
\put(-70,0){\makebox(0,0){$P_1$}}
\put(0,-70){\makebox(0,0){$P_3$}}
\put(70,0){\makebox(0,0){$P_5$}}
\put(0,70){\makebox(0,0){$\cdots$}}
\put(2.5,60){\vector(-1,0){5}}
\ep$$\caption{Type-$n$ Feynman graph. The blobs signify arbitrary $\phi^{2r}$
insertions.}\end{figure}

A diagram with $n$ internal lines is called type-$n$. At one loop, a
type-$n$ graph has $n$ vertices connected to external lines. Since
$2m=\sum_{i=1}^n v_i$,
where $v_i$ is the number of external lines emanating from the $i$th vertex,
the loop integral for the type-$n$ one-loop renormalization of the $2m$ point
function (shown in Figure 1) is
\be
I= \int {d^4 k \over (2 \pi)^4} \label{integrand}
{ \prod_{i=1}^n G^i_{v_i+2} (\ldots, k +Q_{i-1})
\over  (k^2) (k+Q_1)^2 \ldots (k+Q_{n-1})^2 }
\ee
where the upper index on each $G$ labels the relevant vertex as one
goes around the loop.
Here all external momenta are taken to be incoming (signified by
$\ldots$, in the numerator, different for different $G^i$) and we have
defined
$Q_{i}  =  P_i + Q_{i-1} = \sum_{j=1}^{j=i} P_j$,
and $P_i$ is the sum of the external momenta flowing into the
$i$th vertex.
Energy momentum conservation is simply $Q_n = 0$.

For the non-trivial general graph with $n \geq 2$, after combining
denominators and shifting the loop
momenta, we get the integral
\bea
&& (n-1)! \int_0^1 \prod_{j=1}^{n-1} d \alpha_j \int_\ell {d^d \ell \over (
2 \pi)^d} {1 \over D} \\ \nonumber
 &&{\prod_{i=1}^n \mu^{\Delta(i)}  G^i_{v_i +2} (
\ldots , \ell - \sum_{s=1}^{n-1} Q_s \alpha_s + Q_{i-1})}
\label{general}
\eea
where $\alpha_j$ are the Feynman parameters,
\be
D = [\ell^2 - ( \sum_{s=1}^{n-1} Q_s \alpha_s)^2 + (\sum_{r=1}^{n-1} Q_r^2
\alpha_r) -m^2 ]^n
\ee
is the combined denominator, and
$\Delta(i) = { \epsilon \over 2} v_i$
 carries the renormalization scale
 dependence. The ellipsis denote the dependence of each $G^i$ on the
external momenta which play a trivial role everywhere.
 For
the renormalization of a $2m$ point function with $n$ internal lines
$\sum_{i=1}^{i=n} v_i = 2m $,
so summing up the $\mu$ dependence from all the vertices gives the overall
power of $\mu$ appearing as $\mu^{\Delta_{m}}$ with
$\Delta_{m} \equiv \sum_{i=1}^{i=n} v_i( {\epsilon \over 2}) = m \epsilon $.
Now, using the fact that all the $G$'s are analytic, we can write a
symbolic Taylor expansion
 corresponding to the infinite derivative expansion
 \bea
&&\prod_{i=1}^{n} G^i_{v_i+2}(\ldots,l-\sum_{s=1}^{n-1} Q_s\alpha_s + Q_{i-1})
\\ \nonumber
& = &e^{\ell {\partial \over \partial q}} \left[ \prod_{i=1}^{n}
G^i_{v_i+2}(\ldots,q+Q_{i-1}) \right]
\eea
where the $q$ derivatives are evaluated at
${q= - \sum_{s=1}^{n-1} Q_s \alpha_s}$.
Now we just have to do the dimensionally regularized integral
\be
\int{d^{4-\epsilon}\ell
\over(2\pi)^{4-\epsilon}\mu^{-\Delta_m \epsilon}}\,
{e^{\ell{\partial\over\partial q}}
\over\bigl[\ell^2-A^2+i\epsilon\bigr]^n}
\label{drf}
\ee
where
\be
A^2 = ( \sum_{s=1}^{n-1} Q_s \alpha_s)^2 - (\sum_{r=1}^{n-1} Q_r^2
\alpha_r) + m^2 .
\ee

Remembering the analyticity in momenta, we do this by manipulating
the
exponential like a power series, with only even terms in $\ell$ contributing.
Doing the integral, and expanding
the
resultant $\Gamma$ function, with
\be
\Gamma(n-r-2+{\epsilon \over 2}) = {2 \over \epsilon} {(-1)^{2+r-n} (2+r-n)!}
\ee
for $n \leq r+2$,
 we get the pole piece
\bea
&& \sum_{r=n-2}^{\infty} (n-1)! {i \over 8 \pi^2 \mu^{-\Delta_m} \epsilon}
\times \\ \nonumber
&&  { (-1)^{2+r-n}{(-1)}^{n+r}
(A^2)^{2+r-n}
\over {(n-1)! 4^r r! (2+r-n)!}} \,\left[\left({\partial\over\partial
q}\right)^2\right]^r. \label{mainsum}
\eea
Quite nicely, the two factors of $(n-1)!$ cancel (denominator factor
from integration formula and numerator factor from the Feynman trick).
 The sum can then be
written as
\bea
&& { i \over 8 \pi^2 \mu^{-\Delta_m} \epsilon} {\partial^{n-1} \over
\partial x^{n-1}} \\ \nonumber
 && \left[{x \over {(A^2)}^{n-2} } \sum_{r=0}^{\infty}
{(A^2)^{r} x^r
\over  { 4^r r! (r+1)!}} \left[\left({\partial\over\partial
q}\right)^2\right]^r \right],
\eea
with the $x$ derivatives evaluated at $x=1$.
Now, inverting the sum for a Euclidean unit vector, we
may
cast this pole contribution as an integral over a finite
four-dimensional Euclidean angular region to get the $\beta$ function:$^3$

\bea
\lefteqn {\beta_{{G_{2m[v_1,\ldots,v_n]}} (p_1,\ldots,p_{2m-1})} =}
\label{finaleq} \nonumber \\
\nonumber
& & { 1 \over 16 \pi^4
 } { \partial^{n-1} \over \partial x^{n-1} }  \left[ \int d \Omega^4_e
\int_0^1 \prod_{j=1}^{j=n-1} d \alpha_j
{x \over (A^2)^{n-2}} \right. \\ \nonumber
& & \left. \prod_{i=1}^{i=n} G^i_{v_i+2} \left(\ldots,
 \sqrt{x} A e - (\sum_{s=1}^{n-1} Q_s \alpha_s) + Q_{i-1} \right) \right] \\
\nonumber
& & +{\rm cross\atop terms},
\eea
with the $x$ derivatives evaluated at $x=1$.
However, we can do much better, and obtain a more compact expression
in terms of a higher dimensional angular integral. Referring back to the
sum \eqr{mainsum}, we can rewrite
\be
 \sum_{r=n-2}^{\infty}
{ (A^2)^{2+r-n}
\over { 4^r r! (2+r-n)!}} \,\left[\left({\partial\over\partial
q}\right)^2\right]^r
\ee
in terms of $p \equiv r-n+2$ to finally obtain the $1 \over \epsilon$
pole:$^3$
\bea
\lefteqn {{\rm Pole} \, = {i\over 2^{2n-1} \pi^2 \epsilon \mu^{-m\epsilon} }
\left[\left({\partial\over\partial q}\right)^2\right]^{n-2} } \nonumber \\
& &\Biggl[ \int {d \Omega^{2n-2}_e \over \Omega^{2n-2}}
\int_0^1 \prod_{j=1}^{j=n-1} d \alpha_j \\\nonumber
& & \prod_{i=1}^{i=n} G^i_{v_i+2}
\left(\ldots, A e +q
 + Q_{i-1} \right)  \Biggr]
\eea
with the $q$ derivative evaluated at $q=-
(\sum_{s=1}^{n-1} Q_s \alpha_s)$.
This yields the contribution to the $\beta$ function
\it in terms of a $2n-2$ dimensional angular integral over a finite
Euclidean region without reference to any extraneous parameters\rm:
\bea \lefteqn {\beta_{{G_{2m[v_1,\ldots,v_n]}}
(p_1,\ldots,p_{2m-1})} =} \label{finaleq2} \nonumber \\
\nonumber
& & { (n-2)! \over 4^n \pi^{n+1} } \left[ \left({\partial\over\partial
q}\right)^2\right]^{n-2} \\ \nonumber
&& \Biggl[ \int d \Omega^{2n-2}_e
\int_0^1 \prod_{j=1}^{j=n-1} d \alpha_j  \\ \nonumber
& &  \prod_{i=1}^{i=n} G^i_{v_i+2} \left(\ldots,
 A e +q + Q_{i-1} \right) \Biggr] \\
& & +{\rm cross\atop terms} \label{finalres}
\eea
where the $q$ derivative is again evaluated at ${q=- (\sum_{s=1}^{n-1} Q_s
\alpha_s)}$.
Note that a useful check on the result is obtained for $n=2$ which, in the
local limit $G_4 \rightarrow g$ gives the $\beta$ function for local
$\phi^4$ theory: $\beta = {3 g^2 \over 16 \pi^2}$ (after taking account of
a symmetry factor of $1/2$ and a crossing factor of $3$).

We have thus a rather compact expression for the renormalization group
$\beta$ function at one-loop order, for \it any \rm nonlocal coupling
corresponding to `integrating out' a general heavy mass paricle.  The
familiar coefficient for any specific higher derivative operator insertion
may be obtained by expanding both sides of \eqr{finalres} and comparing
powers. Investigation of more direct methods of obtaining our results are
under way.$^4$
The author thanks Prof. H. Georgi for collaboration
on this work, and Dr. B. Schellekens
for invitation to present it at the ICHEP.

\end{document}